\documentclass{jfm}
\usepackage{latexsym}
\usepackage{amsmath}
\usepackage{amssymb}
\usepackage{float}
\usepackage{graphics}
\usepackage{graphicx}
\usepackage{mathrsfs}
\usepackage[toc,page]{appendix}
\usepackage{hyperref}
\hypersetup{
	colorlinks = true,
	urlcolor   = blue,
	citecolor  = blue,
}

\newcommand{\bea}{\begin{eqnarray}}
	\newcommand{\eea}{\end{eqnarray}}

\newcommand{\bc}{\begin{center}}
	\newcommand{\ec}{\end{center}}

\newcommand{\be}{\begin{enumerate}}
	\newcommand{\ee}{\end{enumerate}}
\newcommand{\bt}{\begin{tabbing}}
	\newcommand{\et}{\end{tabbing}}
\newcommand{\btb}{\begin{tabular}}

\title{Weak-electrolyte diffusiophoresis for rigid colloids}
\author{Subrata Majhi\aff{1} \and Huanshu Tan\aff{1}%
	\corresp{\email{tanhs@sustech.edu.cn}}}
\affiliation{
	\aff{1}\parbox[t]{0.95\textwidth}{Multicomponent Fluids Group, Center for Complex Flows and Soft Matter Research, Department of Mechanics and Aerospace Engineering, Southern University of Science and Technology, Shenzhen 518055, Guangdong, P.R. China}
}
\begin{document}
	\maketitle
\begin{abstract}
We develop a model for the diffusiophoresis of a chemically inert, rigid spherical colloid with fixed surface charge in a monovalent weak electrolyte, in which a neutral solute reversibly dissociates into ions. 
A weak far-field gradient is imposed in the neutral-species concentration.
In the fast-reaction limit, local mass action and bulk electroneutrality determine the far-field ionic gradients, while the bulk zero-current condition determines the diffusion-potential gradient.
We solve the coupled Nernst--Planck, Poisson and Stokes equations for arbitrary double-layer thickness, linearising in the gradient strength while retaining the nonlinear Poisson--Boltzmann equilibrium. 
In the Debye--H\"uckel limit, the mobility consists of one half of the matched fully dissociated response and a finite-double-layer correction due to neutral--ion coupling; the correction vanishes in both the H\"uckel and Smoluchowski limits.
Beyond this limit, numerical solutions for the representative systems reveal a branch-selective response as the surface potential magnitude increases.
When the counterion is slower than the co-ion, dissociation--association weakens a retarding concentration-polarisation layer, allowing the mobility to exceed the fully dissociated value.
When the counterion is faster, the response remains close to the one-half scaling set by mass action.
This reaction--polarisation coupling cannot be reproduced by adjusting only the bulk ionic strength, and hence the Debye length, in a fully dissociated model.
\end{abstract}
	
\section{Introduction}
Diffusiophoresis is the migration of a colloidal particle in response to a solute concentration gradient~\citep{anderson1984diffusiophoresis}. 
For a charged particle in an electrolyte, the motion reflects the coupled response of the particle surface and its surrounding electric double layer (EDL). 
Unequal cation and anion diffusivities generate a bulk diffusion potential and hence an electrophoretic contribution, whereas gradients of the excess osmotic pressure within the EDL produce a chemiphoretic contribution~\citep{prieve1984motion,prieve1987diffusiophoresis,velegol2016origins,ault2025physicochemical}.
Finite-double-layer and nonlinear concentration-polarisation effects can produce mobility extrema, while competition between the electrophoretic and chemiphoretic contributions can reverse the migration direction.


Classical electrolyte diffusiophoresis theories generally assume complete dissociation. 
For a fully dissociated binary electrolyte, the imposed salt concentration directly fixes the far-field concentration of each ion, while the zero-current condition determines the diffusion-potential gradient.
A weak electrolyte differs because neutral and ionic species interconvert~\citep{persat2009basic_1,persat2009basic_2,chamberlayne2020effects}.
For the ideal, dilute monovalent reaction $N\rightleftharpoons B^+ + A^-$ considered here, let $n_N^\infty$ be the bulk concentration of the undissociated neutral species and, by bulk electroneutrality, write $n_+^\infty=n_-^\infty=n_I^\infty$ for the concentration of either ion.
For a spatially uniform dissociation equilibrium constant $K_{\mathrm{eq}}$, local mass action in the fast-reaction limit then gives
$n_I^\infty=(K_{\mathrm{eq}}n_N^\infty)^{1/2}$, and hence
$\nabla \ln n_I^\infty
=\tfrac{1}{2}\nabla \ln n_N^\infty$.
Thus, an imposed neutral-species gradient sets both the ionic driving and the equilibrium screening length.
Within the EDL, where electroneutrality does not hold, mass action instead couples the neutral and ionic perturbations.

Weak-electrolyte reactions are known to modify electrokinetic polarisation in electrophoresis.
\citet{baygents1991electrophoresis} showed that mass action tends to oppose disturbances of the diffuse ionic cloud, while \citet{grosse1999electrophoretic} found substantial changes in concentration polarisation and electrophoretic mobility. 
In diffusiophoresis, \citet{timmerhuis2022diffusiophoretic} 
reported smaller particle displacements in carboxylic-acid gradients than predicted by a fully dissociated model, suggesting an effect of partial dissociation. 
These measurements motivate a weak-electrolyte description, but cannot establish whether partial dissociation alone accounts for the reduced motion because pH-dependent surface charging and other solution effects may also contribute. 
An explicit weak-electrolyte theory for rigid-colloid diffusiophoresis at arbitrary EDL thickness is therefore needed.

Here we address this gap for a fixed-charge rigid sphere in a monovalent weak electrolyte in the fast-reaction limit.
The response to a weak neutral-species gradient is obtained for arbitrary double-layer thickness and finite surface potential.
A Debye--H\"uckel solution separates the far-field mass-action scaling from finite-double-layer neutral--ion coupling.
Finite-potential calculations further show that the nonlinear response becomes branch-selective as the magnitude of the surface potential increases.
For the representative system examined, dissociation--association relaxes retarding concentration polarisation sufficiently to overcome the reduced far-field driving on the slow-counterion branch, whereas the fast-counterion branch remains close to the mass-action scaling.
This reaction--polarisation mechanism is absent from fully dissociated models.


\section{Problem formulation}\label{Math_model}

\begin{figure}
\centering
\includegraphics[width=\textwidth]{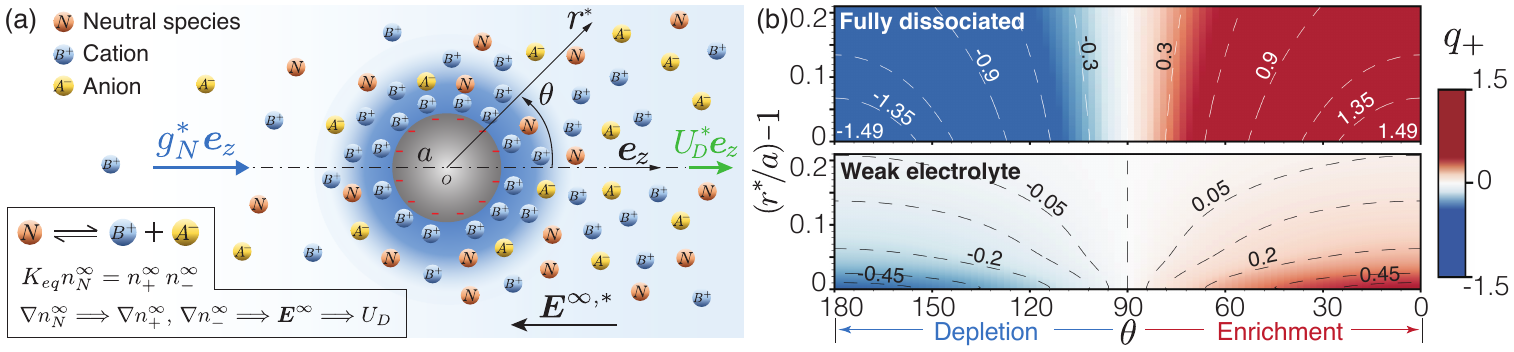}
\caption{%
(a) Schematic of a negatively charged rigid sphere in an imposed neutral-species concentration gradient $g_N^*\boldsymbol{e}_z$, with $N\rightleftharpoons B^++A^-$ in local mass-action equilibrium.
(b) Numerical excess-counterion polarisation, $q_+=[n_+^{(1)}-n_{+,\mathrm{bg}}^{(1)}]\cos\theta$, in the $(r-1,\theta)$ plane for Bis-Tris propane at $\lambda=25$, $\zeta^*=-250~\mathrm{mV}$ and $\gamma=0.01$. 
The top and bottom panels show the matched fully dissociated reference and the weak electrolyte, respectively, on a common colour scale. 
Their imposed background modes are $n_{+,\mathrm{bg}}^{(1)}=f(r)$ and $f(r)/2$, respectively, where $f(r)=r+1/(2r^2)$.}
\label{schematic}
\end{figure}

We consider a rigid spherical colloid of radius $a$ suspended in an ideal, dilute monovalent weak electrolyte. 
The solvent has uniform viscosity $\eta$ and permittivity $\epsilon$, and inertia is neglected. 
As illustrated in figure~\ref{schematic}a, the electrolyte comprises a neutral species $\mathrm{N}$ and its monovalent dissociation products $\mathrm{B}^{+}$ and $\mathrm{A}^{-}$, with $\mathrm{N}\rightleftharpoons \mathrm{B}^{+}+\mathrm{A}^{-}$ and valences $z_+=1$ and $z_-=-1$. 
Their dimensional concentrations are denoted by $n_N^*$, $n_+^*$ and $n_-^*$, respectively; a superscript $*$ denotes a dimensional quantity. 
No supporting electrolyte is present. 
The particle is chemically inert and impermeable. 
Its electrostatic boundary condition is specified by either a spatially uniform surface potential $\zeta^*$ or a spatially uniform surface charge density $\sigma^*$.
A weak, locally uniform far-field gradient $\boldsymbol{g}_N^*=g_N^*\boldsymbol{e}_z$ drives the particle with velocity $U_D^*\boldsymbol{e}_z$.
 We use particle-centred spherical coordinates $(r^*,\theta,\varphi)$, with the polar axis aligned with $\boldsymbol{e}_z$. 
All fields are axisymmetric and $\varphi$-independent.

The dimensional fluxes of the neutral ($N$) and ionic species ($\pm$) are
\begin{align}
    \boldsymbol{J}_N^*
    &=n_N^*\boldsymbol{u}^*
    -\bigl[D_N/(k_BT)\bigr]n_N^*{\nabla}^*\mu_N^*,
    \qquad
    \mu_N^*=\mu_N^{\mathrm{ref},*}+k_BT\ln n_N^*,
    \label{eq:dim_flux_neutral}\\
    \boldsymbol{J}_\pm^*
    &=n_\pm^*\boldsymbol{u}^*
    -\bigl[D_\pm/(k_BT)\bigr]n_\pm^*{\nabla}^*\mu_\pm^*,
    \qquad
    \mu_\pm^*=\mu_\pm^{\mathrm{ref},*}
    +k_BT\ln n_\pm^*+z_\pm e\phi^*.
    \label{eq:dim_flux_ions}
\end{align}
Here $\boldsymbol{u}^*=(u_r^*,u_\theta^*)$ and $\phi^*$ denote the fluid velocity and electric potential. 
The quantities $k_B$, $T$ and $e$ are the Boltzmann constant, absolute temperature and elementary charge, respectively.
The quantities $D_N$ and $D_\pm$ are the species diffusivities, while $\mu_N^*$ and $\mu_\pm^*$ are the corresponding chemical and electrochemical potentials; $\mathrm{ref}$ labels reference values.
The species balances are
\begin{equation}
    {\nabla}^*\cdot\boldsymbol{J}_N^*=-R^*,
    \quad
    {\nabla}^*\cdot\boldsymbol{J}_+^*=R^*,
    \quad 
    {\nabla}^*\cdot\boldsymbol{J}_-^*=R^*,
    \qquad
    R^*=k_d n_N^*-k_a n_+^*n_-^*,
    \label{eq:dim_species_reaction}
\end{equation}
where $k_d$ and $k_a$ are the dissociation and association rate constants.
We consider the fast-reaction limit, in which the chemical relaxation time is short compared with the relevant diffusive transport times.
The reaction therefore remains in local equilibrium,
\begin{equation}
    n_+^*n_-^*=K_{\mathrm{eq}}n_N^*,
    \qquad
    K_{\mathrm{eq}}=k_d/k_a,
    \label{eq:dim_mass_action}
\end{equation}
where $K_{\mathrm{eq}}$ is taken to be spatially uniform. 
Following the auxiliary-balance formulation for weak electrolytes~\citep{baygents1991electrophoresis}, the reaction term can be eliminated by adding the neutral balance to each ionic balance:
\begin{equation}
    {\nabla}^*\cdot
    \left(\boldsymbol{J}_N^*+\boldsymbol{J}_+^*\right)=0,
    \qquad
    {\nabla}^*\cdot
    \left(\boldsymbol{J}_N^*+\boldsymbol{J}_-^*\right)=0.
    \label{eq:dim_auxiliary_balance}
\end{equation}
%
%

The electrostatic and hydrodynamic fields obey the Poisson and Stokes equations
\begin{align}
    -\epsilon{\nabla^*}^2\phi^*
    &=\rho_e^*,
    \qquad
    \rho_e^*=e\left(z_+n_+^*+z_-n_-^*\right),
    \label{eq:dim_poisson}\\
    -\nabla^*p^*
    +\eta{\nabla^*}^2\boldsymbol{u}^*
    -\rho_e^*\nabla^*\phi^*
    &=\boldsymbol{0},
    \qquad
    \nabla^*\cdot\boldsymbol{u}^*=0.
    \label{eq:dim_stokes}
\end{align}
Here $p^*$ is the pressure. 
At $r^*=a$, the no-slip and zero-normal-flux conditions require $\boldsymbol{u}^*=\boldsymbol{0}$, $\boldsymbol{J}_N^*\cdot\boldsymbol{e}_r=0$, and $\boldsymbol{J}_\pm^*\cdot\boldsymbol{e}_r=0$, while the electrostatic condition is either $\phi^*=\zeta^*$ (fixed potential) or $-\epsilon\nabla^*\phi^*\cdot\boldsymbol{e}_r=\sigma^*$ (fixed charge).

Let $n_N^{\infty,*}$ denote the uniform neutral-species concentration in the zero-gradient bulk. 
We define $r=r^*/a$ and the dimensionless gradient strength $G=ag_N^*/n_N^{\infty,*}\ll1$. 
In the electroneutral outer bulk ($n_+^*\sim n_-^*$), local mass action~\eqref{eq:dim_mass_action} gives $n_\pm^*\sim(K_{\mathrm{eq}}n_N^*)^{1/2}$. 
The local bulk ionic reference concentration extrapolated to the particle centre therefore satisfies $(n_I^{\infty,*})^2=K_{\mathrm{eq}}n_N^{\infty,*}$.
Uniform $K_{\mathrm{eq}}$ also gives $\nabla^*\ln n_\pm^*\sim\tfrac12\nabla^*\ln n_N^*$, so the ionic logarithmic gradient is half its neutral-species counterpart.
The $O(G)$ particle-scale far field is thus
\begin{equation}
    n_N^*
    \sim n_N^{\infty,*}\left(1+Gr\cos\theta\right),
    \qquad
    n_\pm^*
    \sim n_I^{\infty,*}
    \left[1+(G/2)r\cos\theta\right],
    \qquad
    1\ll r\ll G^{-1}.
    \label{eq:farfield_dim_concentration}
\end{equation}

The bulk zero-current condition, $\boldsymbol{J}_+^*=\boldsymbol{J}_-^*$, gives in the electroneutral outer bulk
\[
    {\nabla}^*\phi^*
    \sim-(k_BT/e)\beta{\nabla}^*\ln n_\pm^*
    \sim-(k_BT/e)(\beta/2){\nabla}^*\ln n_N^*,
\]
where $\beta=(D_+-D_-)/(D_++D_-)$ is the ionic diffusivity contrast. 
Using \eqref{eq:farfield_dim_concentration} and taking the local outer-bulk potential extrapolated to $z^*=0$ as the reference gives
\begin{equation}
    \phi^*
    \sim-(k_BT/e)(\beta G/2)r\cos\theta,
    \qquad
    1\ll r\ll G^{-1}.
    \label{eq:farfield_dim_potential}
\end{equation}
In the particle-fixed frame, the far-field fluid velocity is
$\boldsymbol{u}^*\sim-U_D^*\boldsymbol{e}_z$.

Together with $r=r^*/a$, we scale electric potential, velocity, pressure, neutral and ionic concentrations, and (electro)chemical potentials by $k_BT/e$, $U_0$, $\eta U_0/a$, $n_N^{\infty,*}$, $n_I^{\infty,*}$ and $k_BT$, respectively, where $U_0=\epsilon(k_BT/e)^2/(\eta a)$. 
Unstarred quantities are dimensionless:
\[
\begin{aligned}
    \zeta&=e\zeta^*/(k_BT),&
    \sigma&=ea\sigma^*/(\epsilon k_BT),&
    U_D&=U_D^*/U_0,\\
    \lambda&=\kappa a,&
    \kappa&=\left[2e^2n_I^{\infty,*}/(\epsilon k_BT)\right]^{1/2},&
    \gamma&=n_I^{\infty,*}/n_N^{\infty,*}.
\end{aligned}
\]
Here $\kappa^{-1}$ is the Debye length and $\gamma$ the ionization ratio.


We introduce the regular expansion $\mathcal{A}(r,\theta)=\mathcal{A}^0(r)+\delta\mathcal{A}(r,\theta)+O(G^2)$, where $\delta\mathcal{A}=O(G)$ and superscript $0$ denotes the spherically symmetric equilibrium state at $G=0$. 
In this state, $\boldsymbol{u}^0=\boldsymbol{0}$ and $\boldsymbol{J}_N^0=\boldsymbol{J}_\pm^0=\boldsymbol{0}$.
The zero-flux conditions make $n_N^0$ and each ionic electrochemical potentials spatially uniform. 
Bulk normalization then gives $n_N^0=1$ and Boltzmann distributions $n_\pm^0(r)=\exp[\mp\phi^0(r)]$, which satisfy local mass action,
$n_+^0n_-^0=n_N^0=1$.
The $O(1)$ part of the dimensionless form of
\eqref{eq:dim_poisson} gives
\begin{equation}
    r^2{\phi^0}''+2r{\phi^0}'
    =(\lambda r)^2\sinh\phi^0,
    \qquad r>1,
    \label{eq:equilibrium_pb}
\end{equation}
with $\phi^0\to0$ as $r\to\infty$ and either $\phi^0(1)=\zeta$ (fixed potential) or $-{\phi^0}'(1)=\sigma$ (fixed charge).
Primes denote differentiation with respect to $r$.
For either boundary condition, we use $\zeta=\phi^0(1)$ to parameterise the equilibrium state.
$\zeta$ is prescribed directly (fixed potential) or is the surface
potential corresponding to $\sigma$ (fixed charge).


The spherical equilibrium state and the dipolar far-field conditions \eqref{eq:farfield_dim_concentration} and \eqref{eq:farfield_dim_potential} restrict the $O(G)$ response to the axisymmetric first spherical harmonic. 
We therefore set $\delta\phi=-G\Phi(r)\cos\theta$, $\delta\mu_N=-G\mathcal{M}_N(r)\cos\theta$ and $\delta\mu_\pm=-z_\pm G\mathcal{M}_\pm(r)\cos\theta$~\citep{baygents1988migration}.
Linearising the chemical potentials in \eqref{eq:dim_flux_neutral}--\eqref{eq:dim_flux_ions} gives $n_N=1-G\mathcal{M}_N\cos\theta$ and
$n_\pm=n_\pm^0[1+z_\pm G(\Phi-\mathcal{M}_\pm)\cos\theta]$.
The dimensionless form of mass action~\eqref{eq:dim_mass_action} then gives $\mathcal{M}_N=\mathcal{M}_+-\mathcal{M}_-$.
The incompressible disturbance flow is represented by $\delta u_r=-(2G/r)\mathcal{U}(r)\cos\theta$ and $\delta u_\theta=G[\mathcal{U}'(r)+\mathcal{U}(r)/r]\sin\theta$~\citep{baygents1991electrophoresis}.


The $O(G)$ Poisson equation obtained from \eqref{eq:dim_poisson} is
\begin{equation}
    \mathcal{L}_1\Phi
    =(\lambda^2/2)\left[
        \exp(-\phi^0)(\Phi-\mathcal{M}_+)
        +\exp(\phi^0)(\Phi-\mathcal{M}_-)
    \right],
    \label{eq:Y_equation}
\end{equation}
where $\mathcal{L}_1q=q''+(2/r)q'-(2/r^2)q$.
The auxiliary ionic balances from \eqref{eq:dim_auxiliary_balance} are
\begin{equation}
	\mathcal{L}_{1}\mathcal{M}_\pm-z_\pm \phi^{0\prime}\mathcal{M}_\pm'+\bigl[\chi_\pm/(z_\pm n_\pm^0)\bigr]\mathcal{L}_{1}\mathcal{M}_N=-2Pe_\pm \phi^{0\prime}{\mathcal{U}}/{r},
    \label{eq:Phi_equation}
\end{equation}
where $Pe_\pm=U_0a/D_\pm$ and $\chi_\pm=D_Nn_N^{\infty,*}/(D_\pm n_I^{\infty,*})=(D_N/D_\pm)\gamma^{-1}$.
Eliminating pressure from the $O(G)$ form of Stokes equation~\eqref{eq:dim_stokes} yields
\begin{subequations}
\begin{align}
    \mathcal{L}_1(\mathcal{L}_1\mathcal{U})
    &=\mathcal{F}_W(r),
    \label{eq:h_equation}\\
    \mathcal{F}_W(r)
    &=-({\lambda^2}/{2r})\phi^{0\prime}(r)
		\left[\exp(-\phi^0)\mathcal{M}_+(r)+\exp(\phi^0)\mathcal{M}_-(r)\right].
    \label{eq:G_general}
\end{align}
\end{subequations}
The explicit $\Phi$-dependence cancels from $\mathcal{F}_W$ because its conservative contribution is absorbed into pressure, leaving $\mathcal{M}_\pm$ to drive vorticity.

At $O(G)$, ionic impermeability and no-slip require $\mathcal{M}_\pm'(1)=0$ and $\mathcal{U}(1)=\mathcal{U}'(1)=0$, respectively. 
The neutral no-flux condition follows automatically from $\mathcal{M}_N=\mathcal{M}_+-\mathcal{M}_-$.
A spatially uniform fixed-potential surface requires
$\Phi(1)=0$, whereas a spatially uniform fixed-charge surface requires
$\Phi'(1)=0$.
Matching the perturbation forms to \eqref{eq:farfield_dim_concentration} and \eqref{eq:farfield_dim_potential} gives $\mathcal{M}_N\sim-r$, $\mathcal{M}_\pm\sim[(\beta\mp1)/2]r$ and $\Phi\sim(\beta/2)r$ as $r\to\infty$. 
The $\mp1/2$ terms reflect the mass-action halving of the ionic concentration gradient, whereas the common $\beta/2$ term arises from the bulk diffusion potential. 
In the particle-fixed frame, the disturbance flow must approach $-U_D\boldsymbol{e}_z$. 
Comparison with the velocity representation gives $\mathcal{U}\sim(m^W/2)r$, so the dimensionless weak-electrolyte diffusiophoretic velocity is $U_D=Gm^W$, where $m^W=2\lim_{r\to\infty}\mathcal{U}(r)/r$ is the dimensionless mobility.


For comparison, we define a matched \textit{fully dissociated} reference with the same $n_I^{\infty,*}$, $D_\pm$ and prescribed surface charge $\sigma$ as the weak electrolyte. 
At fixed $\lambda$, it therefore has the same equilibrium surface potential $\zeta$, P\'eclet numbers $Pe_\pm$ and equilibrium double layer. 
The logarithmic-gradient strength $G$ is imposed directly on the fully dissociated electrolyte, giving its ionic concentration $n_{\pm,S}^*\sim n_I^{\infty,*}(1+Gr\cos\theta)$, whereas mass action halves the weak-electrolyte ionic gradient in \eqref{eq:farfield_dim_concentration}.
The reference contains no neutral species or mass-action constraint, and its mobility is denoted by $m^S$.
It is a controlled same-ionic-strength benchmark rather than a model of the actual weak-acid or weak-base chemistry.

Equations~\eqref{eq:Y_equation}, \eqref{eq:Phi_equation} and \eqref{eq:h_equation}, together with the surface and far-field
conditions, constitute the coupled $O(G)$ problem. 
Numerical solutions were obtained using COMSOL Multiphysics 6.2 with the Coefficient Form PDE interface on $1\leq r\leq R_\infty=30$. 
The equilibrium Poisson--Boltzmann problem was solved first. 
The $O(G)$ transport equations for $\mathcal{M}_\pm$ and the hydrodynamic equations for $(\mathcal{U},\mathcal{Z})$, with $\mathcal{Z}=\mathcal{L}_1\mathcal{U}$, were implemented as two coupled $2\times2$ coefficient-matrix systems.
The potential amplitude $\Phi$ was then recovered from \eqref{eq:Y_equation} to reconstruct the concentration fields. 
The mesh used a maximum radial spacing $\Delta r_{\max}=10^{-3}/\lambda$ over $1\leq r\leq1+2/\lambda$ and was smoothly coarsened to $0.01$ outside this region. 
Fifth-order Lagrange elements and a fully coupled stationary Newton solver with MUMPS were used. 
Increasing $R_\infty$ and refining the mesh changed $m^W$ by less than $0.1\%$.
For acetic acid and Bis-Tris propane, the ionic P\'eclet-number pairs $(Pe_+,Pe_-)$ used in the nonlinear calculations are $(0.056, 0.476)$ and $(0.935, 0.090)$, respectively.

Figure~\ref{schematic}(b) illustrates the resulting finite-potential counterion response. 
For the matched fully dissociated reference, the imposed gradient produces a strong depletion--enrichment dipole that extends beyond the equilibrium double layer. 
The weak-electrolyte response is substantially smaller and more localised. 
Because the two calculations have the same ionic diffusivities and equilibrium double layer, this reduction is consistent with local
dissociation--association buffering counterion depletion and
enrichment. 
Section~\ref{sec:results} quantifies how this change in polarisation affects the mobility. 
The D--H result in section~\ref{sec:DH_analytical} follows by linearising the same equilibrium and $O(G)$ equations for $|\phi^0|\ll1$.


\section{Analytical solution: Debye--H\"uckel limit}
\label{sec:DH_analytical}

To isolate the weak-electrolyte contribution analytically, we consider the Debye--H\"uckel (D--H) regime, $|\phi^0|\ll1$, and retain the mobility through $O(\zeta^2)$. 
Linearising \eqref{eq:equilibrium_pb} gives $(\phi^0)''+(2/r)(\phi^0)'-\lambda^2\phi^0=0$, with solution $\phi^0=\zeta r^{-1}\exp[-\lambda(r-1)]$ satisfying $\phi^0(1)=\zeta$ and $\phi^0\to0$ as $r\to\infty$.
For fixed surface charge, $\sigma=(\lambda+1)\zeta$.

To expose the reaction-induced coupling, we introduce $\mathcal{M}_\Sigma=\mathcal{M}_++\mathcal{M}_-$ and $\mathcal{M}_\Delta=\mathcal{M}_+-\mathcal{M}_-$.
Linearised mass action gives $\mathcal{M}_N=\mathcal{M}_\Delta$, while the far-field conditions give $\mathcal{M}_\Delta\sim-r$ and $\mathcal{M}_\Sigma\sim\beta r$.
Thus, $\mathcal{M}_\Delta$ carries the imposed neutral-species gradient, whereas the leading far-field part of $\mathcal{M}_\Sigma$ encodes the diffusion-potential response. 
Because at the leading order in $\zeta$, $\mathcal{M}_\pm=O(1)$ and $\phi^{0\prime}=O(\zeta)$, \eqref{eq:G_general} and \eqref{eq:h_equation} imply $\mathcal{F}_W, \mathcal{U}=O(\zeta)$. 
For $Pe_\pm=O(1)$, the advective term in \eqref{eq:Phi_equation} therefore changes $\mathcal{M}_\pm$ only at $O(\zeta^2)$ and $\mathcal{F}_W$ only at $O(\zeta^3)$, and may be omitted here. 
Adding and subtracting \eqref{eq:Phi_equation} then give
\begin{equation}
\begin{aligned}
\mathcal{L}_1\mathcal{M}_\Sigma
-\phi^{0\prime}\mathcal{M}_\Delta'
+(\chi_+-\chi_-)\mathcal{L}_1\mathcal{M}_\Delta&=0,\\
(1+\chi_++\chi_-)\mathcal{L}_1\mathcal{M}_\Delta
-\phi^{0\prime}\mathcal{M}_\Sigma'&=0.
\end{aligned}
\label{eq:DH_mode_equations}
\end{equation}

At leading order in $\zeta$, both modes satisfy $\mathcal{L}_1f=0$. 
The solution satisfying $f\sim r$ in the far field and $f'(1)=0$ at the impermeable surface is $f(r)=r+1/(2r^2)$, where $r$ is the imposed uniform-gradient mode and $1/(2r^2)$ is the dipolar disturbance required by surface impermeability. 
Hence, $\mathcal{M}_\Delta^{(0)}=-f$ and $\mathcal{M}_\Sigma^{(0)}=\beta f$.

At $O(\zeta)$, the equilibrium double layer couples the two leading harmonic modes through the common radial source $\phi^{0\prime}f'=\phi^{0\prime}(1-r^{-3})$.
We define $P=O(\zeta)$ by $\mathcal{L}_1P=\phi^{0\prime}(1-r^{-3})$, with $P'(1)=0$ and $P=o(r)$ as $r\to\infty$, so that the correction preserves the surface no-flux condition and the imposed far-field gradient. 
Its Green-function representation is $P=\{rJ_1(r)-f(r)I_1\}/3$, where $I_1=\int_1^\infty\phi^{0\prime}(x)(1-x^{-3})\,\mathrm{d}x$ and $J_1(r)=\int_1^r\phi^{0\prime}(x)(1-x^{-3})(1-x^3/r^3)\,\mathrm{d}x$.
With $C_w=-1+\left[\beta(\chi_--\chi_+)/(1+\chi_++\chi_-)\right]$, whose terms represent the uncoupled finite-EDL response and neutral--ion coupling, respectively, the modes are
\begin{equation}
\mathcal{M}_\Delta=-f+\beta P/\left(1+\chi_++\chi_-\right)+O(\zeta^2),
\quad
\mathcal{M}_\Sigma=\beta f+C_wP+O(\zeta^2).
\label{eq:DH_mode_solutions}
\end{equation}

The hydrodynamic forcing $\mathcal{F}_W$ in the Stokes  equation \eqref{eq:h_equation} depends on the weighted ionic mode $e^{-\phi^0}\mathcal{M}_++e^{\phi^0}\mathcal{M}_-=\mathcal{M}_\Sigma-\phi^0\mathcal{M}_\Delta+O(\zeta^2)$.
Using \eqref{eq:DH_mode_solutions} and $\phi^0,P=O(\zeta)$ gives $\mathcal{M}_\Sigma-\phi^0\mathcal{M}_\Delta
=\beta f+\phi^0f+C_wP+O(\zeta^2)$.
Since $\phi^{0\prime}=O(\zeta)$, \eqref{eq:G_general} reduces to
\begin{equation}
\mathcal{F}_W
=-\frac{\lambda^2\phi^{0\prime}}{2r}
\left(\beta f+\phi^0f+C_wP\right)+O(\zeta^3).
\label{eq:DH_forcing}
\end{equation}
Here $\beta f$, $\phi^0f$, and $C_wP$ represent, respectively, the leading diffusion-potential forcing, quadratic chemiphoretic forcing and neutral--ion-coupled finite-EDL polarisation.

For a no-slip, force-free particle, \eqref{eq:h_equation} gives $m^W=(1/9)\int_1^\infty (1-3r^2+2r^3)\mathcal{F}_W(r)\,\mathrm{d}r$ \citep{ohshima1983approximate}.
Substitution of \eqref{eq:DH_forcing} yields
\begin{equation}
m^W
=\frac{\beta\zeta}{2}\Theta_1(\lambda)
+\frac{\zeta^2}{16}
\left[\Theta_{21}(\lambda)+C_w\Theta_{22}(\lambda)\right]
+O(\zeta^3).
\label{eq:DH_mobility_raw}
\end{equation}
Here $\Theta_1$, $\Theta_{21}$ and $\Theta_{22}$ are the integrated radial responses to the $\beta f$, $\phi^0f$ and $C_wP$ contributions in \eqref{eq:DH_forcing}, respectively.
Their closed forms are $\Theta_1(\lambda)=1+2e^\lambda E_5(\lambda)-5e^\lambda E_7(\lambda)$, where $E_n(x)=\int_1^\infty t^{-n}e^{-xt}\,\mathrm{d}t$ is the generalised exponential integral, and
{\setlength{\jot}{2pt}
\begin{equation}
\begin{aligned}
\Theta_{21}(\lambda)
&=\frac{8\lambda^2}{9}
\left[
\frac{3}{2\lambda}-3e^{2\lambda}E_1(2\lambda)
+\frac{3}{4}e^{2\lambda}
\{E_4(2\lambda)-E_6(2\lambda)\}
\right],\\
\Theta_{22}(\lambda)
&=\Theta_{21}(\lambda)-1
+\frac{8}{3}e^\lambda E_3(\lambda)
-8e^\lambda E_4(\lambda)
-\frac{8}{3}e^\lambda E_5(\lambda)\\
&\quad
+8\Theta_1(\lambda)e^\lambda E_5(\lambda)
+\frac{40}{3}e^\lambda E_6(\lambda)
-\frac{10}{3}e^{2\lambda}E_6(2\lambda)
-\frac{7}{3}e^{2\lambda}E_8(2\lambda).
\end{aligned}
\label{eq:Theta21_Theta22_definition}
\end{equation}}

Let $\Theta_2=\Theta_{21}-\Theta_{22}$. 
Using $\chi_--\chi_+=\beta(\chi_++\chi_-)$, we define $S_\gamma=(\chi_++\chi_-)/(1+\chi_++\chi_-)=\{D_N/D_++D_N/D_-\}/\{\gamma+D_N/D_++D_N/D_-\}$.
Equation \eqref{eq:DH_mobility_raw} then becomes
\begin{equation}
m^W
=\frac{1}{2}
\underbrace{\left[
\beta\zeta\Theta_1(\lambda)
+\frac{\zeta^2}{8}\Theta_2(\lambda)
\right]}_{m^S}
+\frac{\beta^2S_\gamma\zeta^2}{16}\Theta_{22}(\lambda)
+O(\zeta^3).
\label{eq:muD_DH_decomposed}
\end{equation}
Here $m^S$ is the D--H mobility of the matched fully dissociated reference under an imposed ionic logarithmic gradient \citep{keh2000diffusiophoretic,ganguly2024unified}.
The factor $1/2$ reflects the mass-action reduction of the far-field ionic gradient, while the remaining $O(\zeta^2)$ term is the finite-EDL correction due to neutral--ion coupling.

In the H\"uckel ($\lambda\to0$) and Smoluchowski ($\lambda\to\infty$) limits, $(\Theta_1,\Theta_2,\Theta_{22})\to(2/3,0,0)$ and $(1,1,0)$, yielding $m^W\sim\beta\zeta/3$ and $m^W\sim\beta\zeta/2+\zeta^2/16$, respectively.
Within the D--H approximation, the neutral--ion correction therefore vanishes in both limits and is most relevant at intermediate $\lambda$.
We next solve the full $O(G)$ problem numerically beyond the D-H regime.

\section{Results and discussion}\label{sec:results}

Weak ionisation is characterised by the ionisation ratio $\gamma=n_I^{\infty,*}/n_N^{\infty,*}=(K_{\mathrm{eq}}/n_N^{\infty,*})^{1/2}$.
The coupling factor $S_\gamma=(D_N/D_++D_N/D_-)/(\gamma+D_N/D_++D_N/D_-)$ decreases from unity as $\gamma\to0$ ($n_I^{\infty,*}\ll n_N^{\infty,*}$) to zero as $\gamma\to\infty$ ($n_I^{\infty,*}\gg n_N^{\infty,*}$).
Unless stated otherwise, the nonlinear calculations use $\gamma=0.01$, for which the diffusivities considered below give $S_\gamma\simeq0.99$, placing the system in the near-saturated neutral--ion-coupling regime.
All calculations impose a spatially uniform fixed-charge boundary condition, with $\zeta^*$ denoting the corresponding equilibrium surface potential. 
The matched fully dissociated reference has the same $n_I^{\infty,*}$, $D_\pm$ and prescribed surface charge $\sigma$, and hence the same $\lambda$, $\zeta^*$ and equilibrium double layer.
The same dimensionless logarithmic-gradient amplitude $G$ is applied to $n_N^\infty$ in the weak electrolyte and directly to $n_I^\infty$ in the reference.

\begin{figure}
\centering
\includegraphics[width=\textwidth]{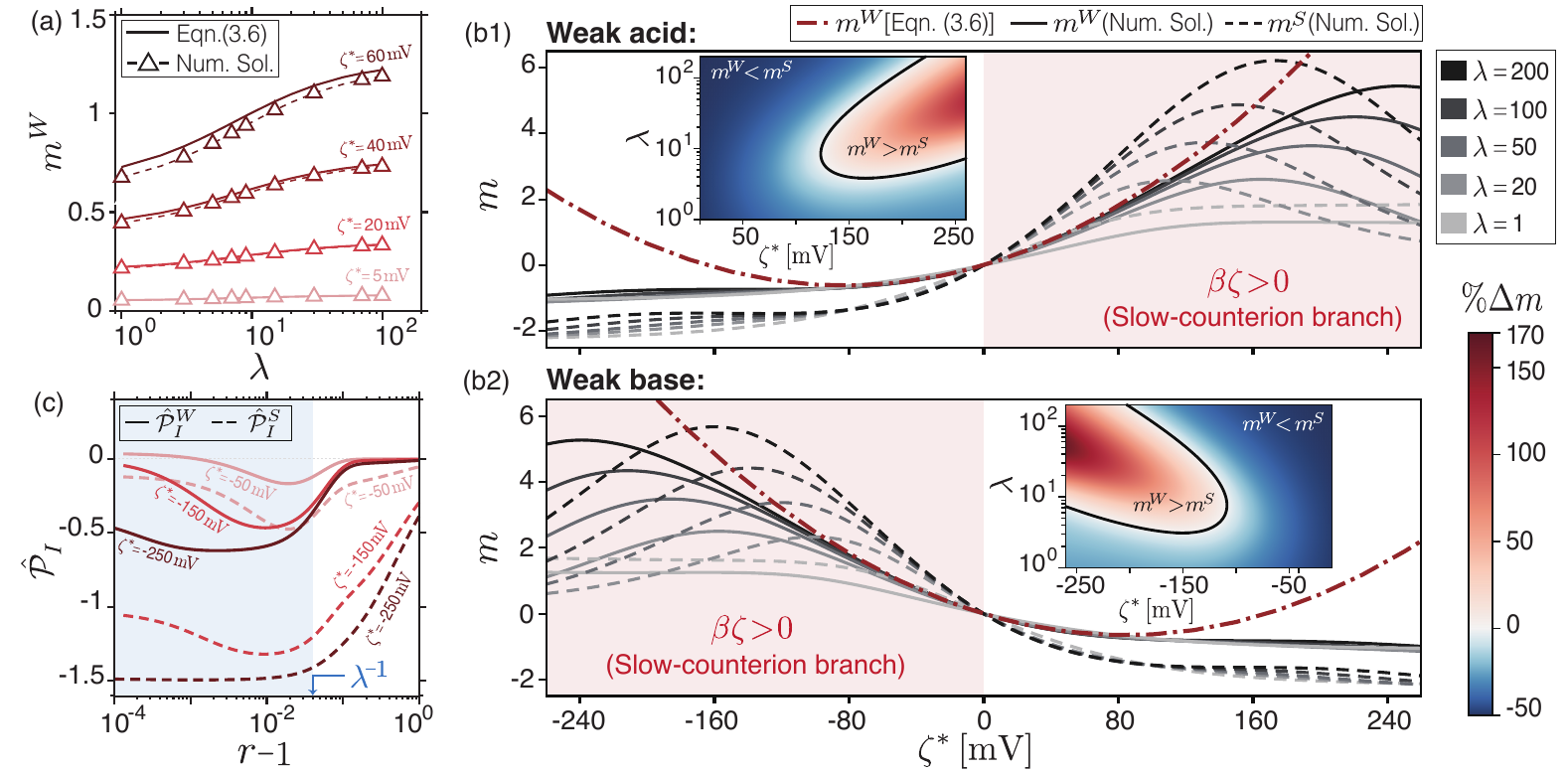}
\caption{%
(a) Acetic-acid mobility $m^W$ from the $O(\zeta^2)$ D--H approximation (Eqn.~\eqref{eq:muD_DH_decomposed}) and nonlinear numerical solutions for $\zeta^*=5,20,40$ and $60~\mathrm{mV}$.
(b1,b2) Numerical mobilities $m^W$ and $m^S$ versus $\zeta^*$ for acetic acid and Bis-Tris propane, respectively, at $\lambda=1,20,50,100$ and $200$; grey levels identify $\lambda$.
The red dash-dotted curve shows the D--H result (Eqn.~\eqref{eq:muD_DH_decomposed}) at $\lambda=200$ and serves as a low-potential reference.
Insets map $\%\Delta m=100(m^W-m^S)/m^S$ on the slow-counterion branches using a common colour scale; the black contour marks $\%\Delta m=0$.
(c) Excess ionic concentration-polarisation mode $\widehat{\mathcal{P}}_I$ for Bis-Tris propane at $\lambda=25$ and $\zeta^*=-50,-150$ and $-250~\mathrm{mV}$.
Solid and dashed curves denote the weak and matched fully dissociated electrolytes, respectively, and the shaded interval marks the nominal Debye-layer region $r-1\leq\lambda^{-1}$.
For acetic acid and Bis-Tris propane, $\beta=0.790$ and $-0.825$, respectively.
}
\label{fig:weak_strong_acetic_btp}
\end{figure}

Figure~\ref{fig:weak_strong_acetic_btp}(a) benchmarks the nonlinear solution against \eqref{eq:muD_DH_decomposed} over $1\leq\lambda\leq100$. 
Across the plotted values of $\lambda$, the maximum relative difference between the numerical and D--H mobilities is below $5\%$ for $\zeta^*\leq40~\mathrm{mV}$ and below $8\%$ at $60~\mathrm{mV}$.
This agreement confirms recovery of the $O(\zeta^2)$ asymptote, while the increasing deviation at $60~\mathrm{mV}$ reflects higher-order finite-potential contributions.

Figures~\ref{fig:weak_strong_acetic_btp}(b1,b2) illustrate the branch selection for acetic acid and Bis-Tris propane, representing a weak acid and weak base, respectively.
Their reactions are 
\[
\mathrm{CH_3COOH\rightleftharpoons H^++CH_3COO^-}\quad \text{and}\quad \mathrm{B+H_2O\rightleftharpoons BH^++OH^-}.
\]
For acetic acid, $(D_N,D_+,D_-)=(1.21,9.31,1.09)$, while for Bis-Tris propane, $(D_N,D_+,D_-)=(0.555,0.555,5.77)$.
The diffusivities are in units of $10^{-9}~\mathrm{m^2\,s^{-1}}$ and give $\beta=0.790$ and $-0.825$, respectively.
At small $|\zeta^*|$, \eqref{eq:muD_DH_decomposed} gives $m^W=m^S/2+O(\zeta^2)$ because mass action halves the far-field ionic logarithmic gradient.
At finite potential, the sign of $\beta\zeta$ distinguishes the slow- and fast-counterion branches.
$\beta\zeta>0$ corresponds to a slow counterion, whereas $\beta\zeta<0$ corresponds to a fast counterion.
Accordingly, acetic acid shows mobility enhancement on the positive-$\zeta^*$ branch, where acetate is the slow counterion, whereas Bis-Tris propane shows mobility enhancement on the negative-$\zeta^*$ branch, where $\mathrm{BH^+}$ is slow.
Additional calculations on the complementary fast-counterion branches give $\%\Delta m$ between approximately $-53\%$ and $-47\%$, consistent with $m^W\simeq m^S/2$.

The suppressed negative-$\zeta^*$ branch of acetic acid is the one relevant to carboxylic-acid experiments on negatively charged polystyrene particles \citep{timmerhuis2022diffusiophoretic}. 
Its near-half-scaled response is qualitatively consistent with the smaller displacements relative to fully dissociated predictions reported there, although those measurements do not isolate mass action from changes in ionic strength, pH and surface charging.

The insets in figures~\ref{fig:weak_strong_acetic_btp}(b1,b2) show $\%\Delta m=100(m^W-m^S)/m^S$.
The black zero contours separate $m^W>m^S$ from $m^W<m^S$ and mark where the mass-action reduction of far-field driving is balanced by the reaction-induced modification of double-layer polarisation.
At $|\zeta^*|=260~\mathrm{mV}$, the strongest sampled enhancement occurs at $\lambda=50$, reaching $123\%$ for acetic acid and $161\%$ for Bis-Tris propane, the latter corresponding to $m^W\simeq2.61m^S$.
Because $|\zeta^*|=260~\mathrm{mV}$ is the endpoint of the sweep, these values are not resolved maxima in $\zeta^*$.
The non-monotonic $\lambda$-dependence is qualitatively consistent with the finite-EDL D--H correction, which vanishes in the H\"uckel and Smoluchowski limits.
These high-potential values are predictions of the point-ion, fixed-charge model, although finite-ion-size and charge-regulation effects may alter their precise magnitudes.

To assess the role of concentration polarisation in the enhancement, we examine the $O(G)$ ionic perturbation introduced in section~\ref{Math_model}.
Writing $n_i=n_i^0+G n_i^{(1)}\cos\theta$ gives $n_i^{(1)}=z_i n_i^0(\Phi-\mathcal{M}_i)$ for $i=\pm$.
We define the mean ionic-concentration perturbation as $\mathcal{P}_I=(n_+^{(1)}+n_-^{(1)})/2$.
Mass action halves the imposed ionic background in the weak electrolyte, so, with $f=r+1/(2r^2)$, the excess perturbations are $\widehat{\mathcal{P}}_I^W=\mathcal{P}_I^W-f/2$ and $\widehat{\mathcal{P}}_I^S=\mathcal{P}_I^S-f$.
They isolate the double-layer-induced perturbation of the mean ionic concentration rather than the charge-density perturbation.

Figure~\ref{fig:weak_strong_acetic_btp}(c) compares $\widehat{\mathcal{P}}_I^W$ (solid) and $\widehat{\mathcal{P}}_I^S$ (dashed) on the slow-counterion branch of Bis-Tris propane at $\lambda=25$.
The shaded interval marks the nominal Debye-layer region $r-1\leq\lambda^{-1}$.
As $|\zeta^*|$ increases, the dashed profiles grow in magnitude and extend well beyond this region, whereas the solid profiles remain smaller and return to zero closer to the surface.
Since $\widehat{\mathcal{P}}_I$ multiplies $\cos\theta$, the smaller solid profiles represent weaker depletion at one pole and weaker enrichment at the other.
The smaller weak-electrolyte profiles are consistent with the neutral species acting as a local reservoir, with dissociation supplying ions in depleted regions and association removing excess ions from enriched regions.
This local mass-action response reduces both the amplitude and radial extent of the double-layer-induced ionic polarisation.

The mobility curves and polarisation profiles reveal two competing weak-electrolyte effects.
Mass action halves the far-field ionic logarithmic gradient, whereas neutral--ion conversion buffers ionic enrichment and depletion and thereby weakens the associated polarisation retardation.
At small $|\zeta^*|$, this retardation is weak, so the loss of far-field driving dominates and $m^W<m^S$.
At larger $|\zeta^*|$ on the slow-counterion branch, the combination of a counterion-rich EDL and unequal ionic transport produces a strong, extended concentration disturbance in the fully dissociated reference.
The associated electrokinetic relaxation opposes the leading phoretic motion, causing $m^S$ to peak and then decrease.
Local mass-action equilibration weakens this disturbance and delays the turnover.
Once the reduction in retardation outweighs the loss of far-field driving, $m^W$ exceeds $m^S$.
For acetic acid at $\lambda=50$, the mobility peak shifts from $\zeta^*\simeq125~\mathrm{mV}$ for $m^S$ to approximately $200~\mathrm{mV}$ for $m^W$, with a similar shift on the negative-$\zeta^*$ branch of Bis-Tris propane.
On the fast-counterion branch, the reference polarisation remains much weaker over the range examined, leaving little retardation for neutral--ion conversion to relax and giving $m^W\simeq m^S/2$.
Thus, $\beta\zeta>0$ selects the branch on which enhancement is possible but does not guarantee it.
Substantial enhancement also requires appreciable $|\beta|S_\gamma$, sufficiently large $|\zeta^*|$, and a finite EDL for which concentration polarisation remains significant.

\section{Conclusion}

We have formulated the diffusiophoresis of a fixed-charge rigid sphere in a monovalent weak electrolyte in the fast-reaction limit. 
When the imposed gradient is specified in the undissociated neutral species, mass action halves the far-field ionic logarithmic gradient. 
The Debye--H\"uckel mobility therefore consists of $m^S/2$ plus the neutral--ion coupling correction $\beta^2 S_\gamma \zeta^2 \Theta_{22}(\lambda)/16$, which vanishes in the H\"uckel and Smoluchowski limits.

The nonlinear solutions show that weak-electrolyte chemistry can either suppress or enhance the mobility relative to the matched fully dissociated reference. 
At low $|\zeta|$,  the loss of far-field ionic driving dominates and $m^W<m^S$.
At larger $|\zeta|$, the response becomes branch-selective. 
For the representative systems examined, a slow counterion ($\beta\zeta>0$) produces a strongly retarding concentration-polarisation layer in the fully dissociated reference. 
Dissociation--association buffers the associated enrichment and depletion, reducing this retardation sufficiently for $m^W$ to exceed $m^S$.
With a fast counterion ($\beta\zeta<0$), the reference polarisation is weaker and the response remains close to $m^W \simeq m^S/2$ over the range examined.

Weak-electrolyte diffusiophoresis therefore cannot, in general, be represented by a fully dissociated model with an adjusted ionic strength alone. 
At large surface potential, quantitative predictions remain conditional on the ideal point-ion Poisson--Boltzmann description. 
More generally, the present theory assumes fixed surface charge, fast local reaction equilibrium and a weak gradient imposed in the neutral species. 
Extensions incorporating pH-dependent charge regulation, finite-rate
chemistry and gradients specified in total solute concentration would permit closer comparison with experiments. \\

H.T. acknowledges support from the National Natural Science Foundation of China (Grant Nos. 12588301, 12472271) and the Guangdong Basic and Applied Basic Research Foundation (Grant No. 2024A1515010614).

Declaration of Interests: The authors report no conflict of interest.

During the preparation of this manuscript, the authors used OpenAI Codex to assist with language editing, manuscript organisation, and checks of the internal consistency of the mathematical derivation. 
All AI-assisted output was critically reviewed and independently verified by the authors, who take full responsibility for the content. 
No AI tool was used to generate research data, numerical results, or manuscript figures.

\bibliographystyle{jfm_v2}
\bibliography{reference}
\end{document}